\documentclass{aa}
\usepackage{txfonts}
\usepackage{graphicx}
\usepackage{setspace}
\usepackage{natbib}
\usepackage{longtable}
\usepackage{amssymb}
\bibpunct{(}{)}{;}{a}{}{,}

\begin{document}

\newcommand{\FeII}{[\ion{Fe}{ii}]}
\newcommand{\TiII}{[\ion{Ti}{ii}]}
\newcommand{\SII}{[\ion{S}{ii}]}
\newcommand{\OI}{[\ion{O}{i}]}
\newcommand{\OIp}{\ion{O}{i}}
\newcommand{\PII}{[\ion{P}{ii}]}
\newcommand{\NI}{[\ion{N}{i}]}
\newcommand{\NII}{[\ion{N}{ii}]}
\newcommand{\NIp}{\ion{N}{i}}
\newcommand{\NiII}{[\ion{Ni}{ii}]}
\newcommand{\CaIIp}{\ion{Ca}{ii}}
\newcommand{\PI}{[\ion{P}{i}]}
\newcommand{\CIp}{\ion{C}{i}}
\newcommand{\HeI}{\ion{He}{i}}
\newcommand{\MgIp}{\ion{Mg}{i}}
\newcommand{\MgIIp}{\ion{Mg}{ii}}
\newcommand{\NaI}{\ion{Na}{i}}
\newcommand{\HI}{\ion{H}{i}}
\newcommand{\brg}{Br\,$\gamma$}
\newcommand{\pab}{Pa\,$\beta$}

\newcommand{\macc}{$\dot{M}_{acc}$}
\newcommand{\lacc}{L$_{acc}$}
\newcommand{\lbol}{L$_{bol}$}
\newcommand{\mjet}{$\dot{M}_{jet}$}
\newcommand{\mh}{$\dot{M}_{H_2}$}
\newcommand{\Ne}{n$_e$}
\newcommand{\h}{H$_2$}
\newcommand{\kms}{km\,s$^{-1}$}
\newcommand{\um}{$\mu$m}
\newcommand{\lam}{$\lambda$}
\newcommand{\msyr}{M$_{\odot}$\,yr$^{-1}$}
\newcommand{\Av}{A$_V$}
\newcommand{\msun}{M$_{\odot}$}
\newcommand{\lsun}{L$_{\odot}$}
\newcommand{\cm}{cm$^{-3}$}

\newcommand{\bet}{$\beta$}
\newcommand{\alfa}{$\alpha$}

\hyphenation{mo-le-cu-lar pre-vious e-vi-den-ce di-ffe-rent pa-ra-me-ters ex-ten-ding a-vai-la-ble}

\title{The nature of the embedded intermediate-mass T Tauri star DK Cha\thanks{Based on observations collected at the European Southern 
Observatory, La Silla and Paranal, Chile (ESO program 082.C-0264(A) and 084.C-0308(A)).}}
\author{Rebeca Garcia Lopez \inst{1,2,3} \and Brunella Nisini \inst{2} \and Simone Antoniucci \inst{2} \and Alessio Caratti o Garatti \inst{1} \and Dario Lorenzetti \inst{2} \and Teresa Giannini \inst{2} \and Jochen Eisl\"{o}ffel \inst{4} \and Tom Ray \inst{1}}

\offprints{R. Garcia Lopez, \email{rgarcia@mpifr.de}}

\institute{Dublin Institute for Advanced Studies, School of Cosmic Physics, 31 Fitzwilliam Place, Dublin 2, Ireland \and INAF-Osservatorio Astronomico di Roma, Via di Frascati 33, I-00040 Monteporzio Catone,
Italy \and Max-Planck-Institut f\"{u}r Radioastronomie, Auf dem H\"{u}gel 69, D-53121 Bonn, Germany \and Th\"{u}ringer Landessternwarte Tautenburg, Sternwarte 5, 07778 Tautenburg, Germany}

%----------------------------------------------------------------------
%
\date{Received date; Accepted date}
%
%----------------------------------------------------------------------
%
%
\titlerunning{Dk Cha}
\authorrunning{Garcia Lopez, R. et al.}

\abstract
{Most of our knowledge about star formation is based on studies of low-mass stars, whereas very little is known about the properties of the circumstellar material around young and embedded intermediate-mass T Tauri stars (IMTTSs) mostly because they are rare, typically more distant than their lower mass counterparts, and their nearby circumstellar surroundings are usually hidden from us.}
{We present an analysis of the excitation and accretion properties of the young IMTTS DK Cha. The nearly face-on configuration of this source allows us to have direct access to the star-disk system through the excavated envelope and outflow cavity. }
{Based on low-resolution optical and infrared spectroscopy obtained with SofI and EFOSC2 on the NTT we derive the spectrum of DK Cha from $\sim$0.6\,\um\ to $\sim$2.5\,\um. From the detected lines we probe the conditions of the gas that emits the \HI\ IR emission lines and obtain insights into the origin of the other permitted emission lines. In addition, we derive the mass accretion rate (\macc) from the relationships that connect the luminosity of the \brg\ and \pab\ lines with the accretion luminosity (\lacc).}
{The observed optical/IR spectrum is extremely rich in forbidden and permitted atomic and molecular emission lines, which makes this source similar to very active low-mass T Tauri stars. Some of the permitted emission lines are identified as being excited by fluorescence. 
%From the color excesses we have derived a visual extinction of 11.2$\pm$1\,mag towards the source. This value is in agreement with that derived %form the silicate absorption feature at 10\,\um.
We derive Brackett decrements and compare them with different excitation mechanisms. The Pa$\beta$/Br$\gamma$ ratio is consistent with optically thick emission in LTE at a temperature of $\sim$ 3500 K, originated from a compact region of $\sim$ 5 R$_\odot$ in size: but the line opacity decreases in the Br lines for high quantum numbers n$_{up}$. A good fit to the data is obtained assuming an expanding gas in LTE, with an electron density at the wind base of  $\sim$10$^{13}$\,\cm. In addition, we find that the observed Brackett ratios are very similar to those reported in previous studies of low-mass CTTSs and Class I sources, indicating that these ratios are not dependent on masses and ages. 
Finally, L$_{acc}\sim$9\,\lsun\ and $\dot{M}_{acc}\sim$3$\times$10$^{-7}$\,\msyr\ values were found. When comparing the derived \macc\ value with that found in Class I and IMTTSs of roughly the same mass, we found that \macc\ in DK Cha is lower than that found in Class I sources but higher than that found in IMTTSs. This agrees with DK Cha being in an evolutionary transition phase between a Class I and II source.
%The derived \macc\ value is lower than that found in Class I sources of the same mass but higher than that found in intermediate mass TTauri stars. This agrees with DK Cha being in an evolutionary transition phase between a Class I and II source.
%, similar to those found in typical Herbig Ae stars. 
%This indicates that DK Cha is in its final accretion stage, with an L$_{acc}$/L$_{bol}$ less than 0.5, or that most of the matter is accumulated in outbursts of accretion   that could be responsible for the high variability of the object.
}
{}

\keywords{stars: formation -- stars:circumstellar matter -- stars: pre-main sequence -- ISM:individual objects: DK Cha, IRAS 12496-7650 -- Infrared: ISM}

\maketitle

%
%------------------------------------------------------------------------- 
%---------------------------------------------------------------------------

\section{Introduction}

Intermediate-mass TTauri stars are the young precursors of Herbig AeBe stars with spectral types ranging from K to late F, and ranging from 1\,\msun\ to 5\,\msun\ (e.g. \citealt{calvet04}). Their disks dissipate in $\sim$5\,Myr, i.e., more quickly than their low-mass classical TTauri (CTT) counterparts, which makes it very difficult to study them in the early evolutionary phase.
Besides, the complex structure of the emission region surrounding protostars makes it more difficult to probe the disk accretion properties, a crucial step towards understanding disk evolution in intermediate- mass stars.

In the nearby surroundings of protostars one does indeed expect emission from the heated dense envelope, the disk, the jet and the UV-heated and shocked cavity walls, excavated by the outflow.
Information about the different phenomena involving the emission from each of the components of this system can only be probed through spectroscopic studies that allow us to obtain quantitative information on processes that take place in spatially unresolved regions (such as the accretion through funnel flows and the ejection of matter from the inner disk). For instance, the jet properties can be studied through forbidden emission lines, while the accretion properties are usually probed through \HI\ transitions.
In this context, infrared spectroscopy turns out to be a very useful tool to study embedded YSOs where the high extinction usually prevents us from observing them through the standard optical tracers.
However, in a small number of cases, when the object is observed face-on, we can have direct access to the star-disk system through the outflow cavity and excavated envelope even at shorter wavelengths.

%%%%%%%%%%%%%%%%%%%%%%%		spectra		%%%%%%%%%%%%%%%%%%%%%%%%%%%%
\begin{figure*}[!t]
\resizebox{\textwidth}{!}{\includegraphics{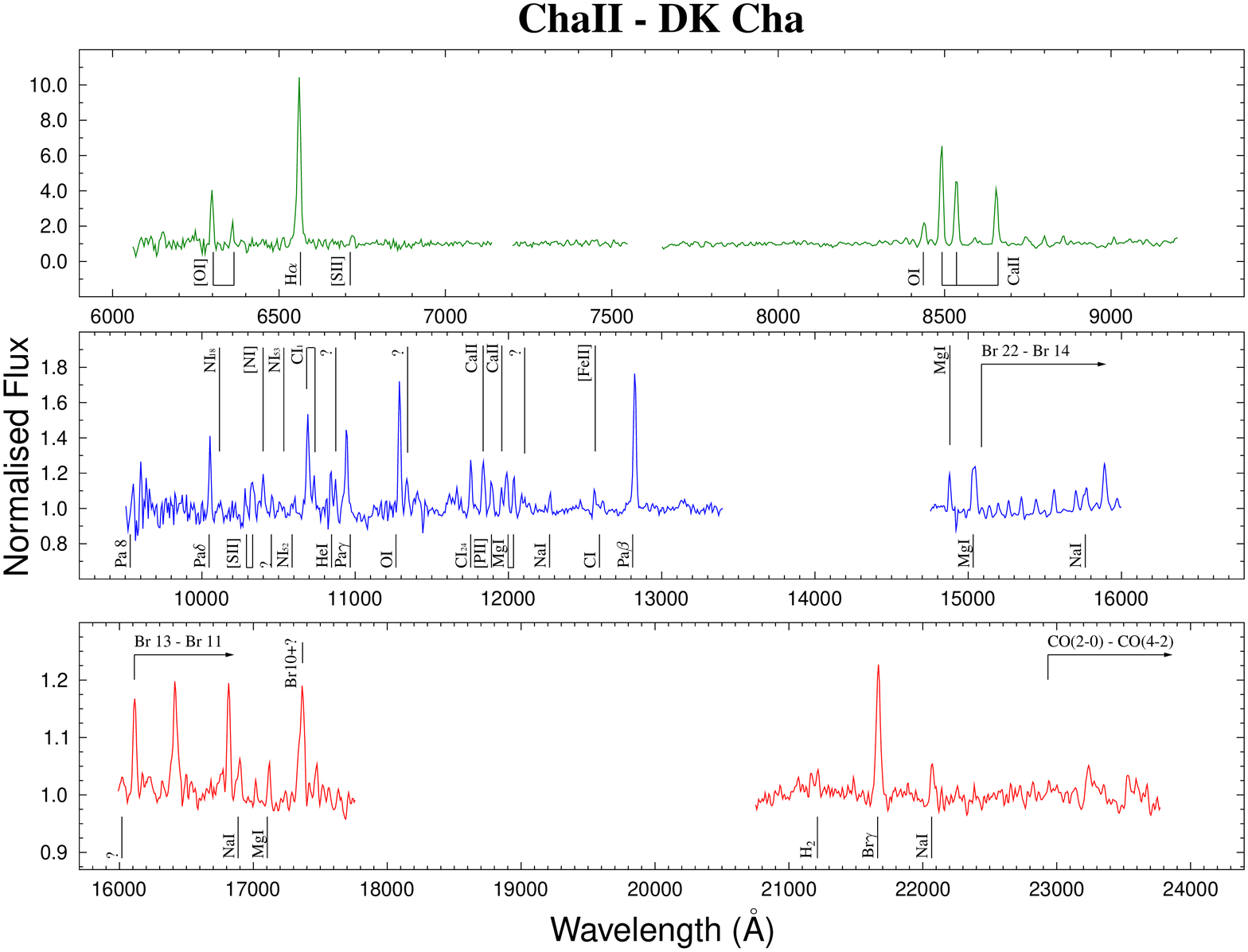}}
\caption{Normalised to the continuum EFOSC2+SofI spectra from 0.6 to 2.4\,\um\ of the PMS star DK\,Cha. The strongest lines are identified.}
  \label{fig:spectra}
\end{figure*}
%%%%%%%%%%%%%%%%%%%%%%%%%%%%%%%%%%%%%%%%%%%%%%%%%%%%%%%%%%%%%%%%%%%%%%%%%%%%%

The IMTTS  DK Cha (M$_*$=2\,M$_\odot$, L$_{bol}$=29.4\,L$_\odot$) is one of these rare examples where, thanks to its nearly face-on configuration (i$\lesssim$18\degr), the star-disk system can be studied through optical/infrared tracers \citep{vankempen09,vankempen10,spezzi08,hughes91}. DK Cha has been identified as a young object still surrounded by its envelope, as shown by the large amount of warm molecular gas detected on-source and the presence of a strong molecular outflow \citep{vankempen09,vankempen06}. In addition, its spectral energy distribution (SED) corresponds to that of an YSO in a transition phase from a Class I to a Class II source \citep{vankempen10,spezzi08}. 
These characteristics together with its nearby location (d$\sim$178\,pc) make DK Cha an ideal candidate to probe the first stages of the formation of Herbig Ae stars.

Most of our knowledge about this source is based on far-IR and sub-millimeter observations \citep[e.g.,][]{vankempen06,vankempen10,giannini99}. Optical spectroscopic studies of DK Cha have been made by \cite{hughes91}. In that work the first extensive study of the optical and infrared photometric variability of DK Cha was also reported, pointing out the highly variable nature of this source.
Here, we will present an optical/IR spectroscopic study of DK Cha aimed at obtaining insights into the characteristics of its circumstellar material and accretion properties. This study was conducted as part of the POISSON (Protostellar Objects IR-optical Spectral Survey On NTT) project, whose results will be presented in a series of papers on the properties of YSOs belonging to different clouds \citep[e.g. ChaI-II regions and L1641][]{simone11, aleL1641}.

In Sect.\,\ref{sect:observations} the observations and the data reduction will be described, while in Sect.\,\ref{sect:lines} the first complete spectrum of DK\,Cha from 0.6 to 2.4\,\um\ will be presented. In Sect.\,\ref{sect:results} we show the analysis and discuss our observations. Finally,  we present the conclusions in Sect.\,\ref{sect:conclusions}.

\section{Observations and data reduction}
\label{sect:observations}
%%%%%%%%%%%%%%%%%%%%%%%%%%%%%%%%%%%%%%%%%%%%%%%%%%%%%%%%%%%%%%%%%%%%%%%%%%%%%
\begin{table}
\begin{minipage}[!th]{\columnwidth}
\caption{Detected emission lines in DKCha.}
\label{tab:lines}
\centering
\renewcommand{\footnoterule}{}  % to avoid a line before footnotes
\begin{tabular}{c c c c c}

\hline \hline
Line id. & $\lambda$ & EW &  F & $\Delta$F \\
 ~       &  (\um)    & (\AA)	& \multicolumn{2}{c}{(10$^{-14}$ erg cm$^{-2}$ s$^{-1}$)} \\
\hline 

\OI\,$^1$D$_2 - ^3$P$_2$ 		& 0.630 & -30.3  & 1.59 & 0.06 \\
\OI\,$^1$D$_2 - ^3$P$_1$ 		& 0.636 & -19.4  & 0.83 & 0.10 \\
H$\alpha$	     			& 0.656 & -135.8 & 9.28 & 0.18  \\
\SII\,$^2$D$_{5/2} - ^4$S$_{3/2}$	& 0.672 & -8.5   & 0.70 & 0.10 \\
\OIp\,$^3$P$ - ^3$S$^{0\, (m)}$
%\footnote{Multiplet: $2$s$^2$.$2$p$^3$.($^4$S$^o$).$3$s-$2$s$^2$.$2$p$^3$.($^4$S$^o$).3p; J$_i=1$, J$_k=0,2,1$}					
			& 0.845 & -17.9	 & 5.44 & 0.47 \\

%Pa\,17				& 0.847 & 2.85E-14 & 1.01E-14 \\
\CaIIp\,$^2$P$_{3/2}-^2$D$_{3/2}$	& 0.850 & -75.2  & 22.7 & 0.41  \\
\CaIIp\,$^2$P$_{3/2}-^2$D$_{5/2}$	& 0.854 & -53.9	 & 16.9 & 0.01  \\
\CaIIp\,$^2$P$_{1/2}-^2$D$_{3/2}$	& 0.866 & -38.4	 & 14.3 & 0.18   \\
Pa\,12					& 0.875 & -8.1	 & 3.04 & 0.22	\\
\PI\,$^2$D$^0-^4$S$^{0\,(m)}$ 
%\footnote{Multiplet: $3$s$^2$.$3$p$^3$-$3$s$^2$.$3$p$^3$; J$_i=3/2$, J$_k=5/2, 3/2$}
					 & 0.879 & -6.0  & 2.15 & 0.12	 \\
%\PI\,$^2$D$^0_{3/2}-^4$S$^{0*}_{3/2}$    & 0.880 &          &           \\
%Pa\,11					& 0.886 & -7.1   & 2.61 & 0.39	 \\
%Pa\,10					& 0.902 & -2.7	& 1.06 & 0.04	 \\
%Pa\,8					& 0.955 &  	&      & 	 \\
Pa\,$\delta$				& 1.005 & -8.2 	& 4.65 & 0.42	 \\
\NIp\,$^4$F$_{3/2}-^4$D$^{0\,(m)}_{3/2}$
%\footnote{Fluorescent line, multiplet 18 (see, \citealt{walmsley00})}
 					& 1.013 &  	&      &   \\
\SII\,$^2$P$_{3/2}-^2$D$_{3/2}$		& 1.029 & -3.4	& 2.21 & 0.37	 \\
\SII\,$^2$P$_{1/2}-^2$D$_{3/2}$ 	& 1.034 & -7.2 	& 5.02 & 0.67	 \\
\NI\,$^2$P$^0-^2$D$^{0\,(m)}$
% \footnote{Multiplet:$2$s$^2$.$2$p$^3$-$2$s$^2$.$2$p$^3$; J$_i$=$5/2,3/2$, J$_k$=$3/2,1/2$}
					& 1.040 & -3.6 	& 2.80 & 0.39	 \\
%\NI\,$^2$P$^0_{1/2}-^2$D$_{5/2}^{0*}$&  ~    &   ~	   &    ~     \\
\NiII\,$^4$P$_{5/2}-^2$F$_{7/2}$	 & 1.046 & -2.4 & 1.89 & 0.52	\\
\NIp\,$^4$D$ -^4$P$^{0\,(m)}$
% \footnote{Fluorescent line, multiplet 53 (see, \citealt{walmsley00})}
					 & 1.051 & -2.8 & 2.32 & 0.60	 \\
\NIp\,$^4$P$-^4$P$^{0\,(m)}$
%\footnote{Fluorescent line, multiplet: 52}
					& 1.062 & -3.3  & 2.83 & 0.41	 \\
\CIp\,$^3$D$-^3$P$^{0\,(m)}$
% \footnote{Recombination line; multiplet 1 (see, \citealt{walmsley00}); J$_i$=1,0, J$_k$=2,1}	
					& 1.069 & -10.5 & 10.2 & 0.35	 \\
\CIp\,$^3$D$_2-^3$P$^{0\,(m)}_2$
%\footnote{Recombination line; multiplet 1 (see, \citealt{walmsley00})}
					& 1.073 & -2.1 & 2.16 & 0.21	 \\
\HeI\,					& 1.083 & -3.0 & 3.18 & 0.20	 \\
Pa\,$\gamma$				& 1.094 & -11.6 & 12.6 & 1.33	\\
\OIp\,$^3$P$-^3$D$^{0\,(m)}$
%\footnote{Fluorescent line; multiplet 18.02 (see, \citealt{walmsley00})}
					& 1.129 & -15.1 & 20.0 & 0.78	 \\
\CIp$^{(m)}$
%\footnote{Recombination line; multiplet 24 (see, \citealt{walmsley00}); J$_i$=1,3,2; J$_k$=2,4,3 }
					& 1.175 & -5.8 & 9.41 & 0.33	 \\
\CaIIp\,$^2$P$_{3/2}-^2$S$_{1/2}$	& 1.184 & -7.7 & 12.8 & 0.67	 \\
\PII\,$^3$P$_2$-$^1$D$_2$		& 1.188 & -4.8 & 8.29 & 0.68	 \\
\CaIIp\,$^2$P$_{1/2}-^2$S$_{1/2}$	& 1.195 & -1.9 & 3.37 & 0.37	 \\
\MgIp\,$^1$P$^0_1-^1$S$_0$		& 1.198 & -4.6 & 8.47 & 1.54	 \\
\MgIp\,$^1$P$^0_1-^1$D$_2$		& 1.203 & -3.5 & 6.39 & 0.11	\\

\hline

\end{tabular}
\tablefoot{ (m) indicates lines belonging to multiplets.
	    Observed fluxes, not corrected for extinction.}
\end{minipage}
\end{table}
%%%%%%%%%%%%%%%%%%%%%%%%%%%%%%%%%%%%%%%%%%%%%%%%%%%%%%%%%%%%%%%%%%%%%%%%%%%%

\addtocounter{table}{-1}
%%%%%%%%%%%%%%%%%%%%%%%%%%%%%%%%%%%%%%%%%%%%%%%%%%%%%%%%%%%%%%%%%%%%%%%%%%
\begin{table}[!th]

%\begin{minipage}[t]{\columnwidth}
\caption{Continued.} 
%\label{tab:lines}
\centering
\begin{tabular}{c c c c c}
\hline \hline

Line id. & $\lambda$ & EW & F & $\Delta$F \\
 ~       &  (\um)    & (\AA) & \multicolumn{2}{c}{(10$^{-13}$ erg cm$^{-2}$ s$^{-1}$)} \\
\hline

\NaI\,$^2$P$^0-^2$S$^{(m)}$ 
							& 1.227 & -1.9 &  0.39 & 0.03	\\
%\NaI\,$^2$P$^0_{1/2}-^2$S$_{1/2}^*$ &    ~  &   ~      &   ~      \\
\FeII\,a$^4$D$_{7/2}$-a$^6$D$_{9/2}^{(blend)}$		& 1.257 & -1.1 & 0.27 & 0.02	\\
\CIp\,$^3$P$^0-^3$P$^{(blend;\,m)}$	& 1.261 & -0.5 & 0.12 & 0.04	\\
Pa\,$\beta$						& 1.282 & -16.9 & 4.43 & 0.09	 \\
\MgIp\,$^1$S$^0-^1$P$^{0\,(m)}_0$     			& 1.488 & -3.7 & 1.99 & 0.11	 \\
\MgIp\,$^3$P$^0-^3$S$^{\,(m)}$ 
							& 1.504 & -10.5 & 5.67 & 0.33	 \\
%Br\,22							& 1.509 &       &      & 	 \\
%Br\,21							& 1.514 &       &      & 	 \\
Br\,20							& 1.519 & -3.0 & 1.65 & 0.20 	 \\
Br\,19							& 1.526 & -1.5 & 0.84 & 0.24	 \\
Br\,18							& 1.534 & -1.9 & 1.09 & 0.19	 \\
Br\,17							& 1.544 & -2.0 & 1.14 & 0.30	 \\
Br\,16							& 1.556 & -3.1 & 1.84 & 0.11	 \\
Br\,15							& 1.570 & -2.7 & 2.02 & 0.12	 \\
\NaI\,$^2$S$-^2$P$^{0\,(m)}$  
							& 1.576 & -4.1 & 2.65 & 0.25	 \\
Br\,14							& 1.588 & -7.7 & 5.04 & 0.25	 \\
\FeII\,a$^4$D$_{3/2}$-a$^4$F$_{7/2}$			& 1.600 & -1.1 & 0.72 & 0.23	 \\
%?							& 1.602 & -1.3 & 0.87 & 0.53	\\
Br\,13							& 1.611 & -5.5 & 3.34 & 0.23	 \\
Br\,12$^{(blend)}$					& 1.641 & -9.0 & 6.62 & 0.51	 \\
\FeII\,a$^4$D$_{7/2}$-a$^4$F$_{9/2}^{(blend)}$		& 1.644 &  ~   &      & ~	\\
Br\,11							& 1.681 & -5.8 & 4.69 & 0.20	 \\
\NaI\,$^2$D$-^2$P$^{0\,(m)}$
							& 1.689 & -2.9 & 2.41	& 0.29	 \\
\MgIp\,$^1$P$^0_1$-$^1$S$_0$				& 1.711 & -2.0 & 1.77 & 0.20	\\
Br\,10$^{(blended with ?)}$					& 1.737 & -7.7 & 8.35 	& 0.66	 \\
H$_2$1-0S(1)						& 2.122 & -0.7 & 1.14 	& 0.31	\\
\brg\							& 2.166 & -8.8 & 15.0	& 0.58	 \\
\NaI\,$^2$P$^0_{3/2}-^2$S$_{1/2}^{(blend)}$		& 2.206 & -2.2 & 3.87 	& 0.54	\\
\NaI\,$^2$P$^0_{1/2}-^2$S$_{1/2}^{(blend)}$		& 2.209 & ~	& ~	 \\
CO(2-0)							& 2.293	& -0.6 & 1.17	& 0.24 	\\
CO(3-1)							& 2.323 & -2.7 & 5.20	& 0.60	\\
CO(4-2)							& 2.353 & -2.2 & 4.33	& 0.64	\\	

\hline

\end{tabular}
\tablefoot{ (m) indicates lines belonging to multiplets.
Observed fluxes, not corrected for extinction.}
%\tablefoottext{a}{Multiplet: 2p$^6$.5s-2p$^6$.34p; J$_i$=1/2, J$_k$=3/2, 1/2} 
%\tablefoottext{b}{recombination line?} 
%\tablefoottext{c}{Multiplet: 3s.4s-3s.4p; J$_i$=1, J$_k$=2,1,0} 
%\tablefoottext{d}{Multiplet: 2p6.5p-2p6.43s; J$_i$=1/2 J$_k$=1/2, 3/2} 
%\tablefoottext{e}{Multiplet: 2p$^6$.5p-2p$^6$.15d; J$_i$=3/2-5/2, J$_k$=1/2, 3/2} 

%\footnote{Multiplet: 2p$^6$.5s-2p$^6$.34p; J$_i$=1/2, J$_k$=3/2, 1/2 }
%\footnote{recombination line?}
%\footnote{Multiplet: 3s.4s-3s.4p; J$_i$=1, J$_k$=2,1,0}
%\footnote{Multiplet: 2p6.5p-2p6.43s; J$_i$=1/2 J$_k$=1/2, 3/2}
%\footnote{Multiplet: 2p$^6$.5p-2p$^6$.15d; J$_i$=3/2-5/2, J$_k$=1/2, 3/2}
%\end{minipage}
\end{table}
%%%%%%%%%%%%%%%%%%%%%%%%%%%%%%%%%%%%%%%%%%%%%%%%%%%%%%%%%%%%%%%%%%%%%%%%%%%%%
 
As part of the POISSON project, we observed DK\,Cha in the wavelength range from $\sim$0.6\,\um\ to $\sim$2.5\,\um\ using the optical and infrared spectrographs EFOSC2 and SofI at the ESO-NTT telescope \citep{efosc,sofi}. The similar spectral resolution and the selection of comparable slit widths for both instruments (R$\sim$700-900 and slits widths of 0\farcs7 and 0\farcs6 for EFOSC2 and SOFI, respectively) allowed us to acquire optical and infrared spectra of homogeneous characteristics.

The EFOSC2 spectrum (hereafter \textquotedblleft optical spectrum\textquotedblright) was acquired with grism\,16 and roughly ranges from 0.6\,\um\ to 1.0\,\um.
The infrared SOFI spectra were taken using both the blue- (GB: 0.95\,\um -1.64\,\um) and red-grism (GR: 1.53\,\um-2.52\,\um).
The optical and near-IR (NIR) data were acquired close in time on 12$^{th}$ and 16$^{th}$ February 2009, respectively.
%, allowing us to avoid 
%relevant flux line variations in the different spectral segments due to stellar variability.
The total integration times on DK\,Cha were 300\,s for the optical and 60\,s and 20\,s for the blue and red IR grim spectra. These were acquired through standard A-B nodding.

For data reduction standard IRAF\footnote{IRAF (Image Reduction and Analysis Facility) is distributed by the National Optical Astronomy Observatories, which are operated by AURA, Inc., cooperative agreement with the National Science Foundation.} tasks were used. Wavelength calibrations were performed using argon-xenon lamps. 
The NIR spectra were corrected for the atmospheric spectral response by dividing them by the spectrum of a suitable standard star. The three overlapping segments were then joined to obtain a single optical/IR spectrum of DK Cha. 
Unfortunately, during the two observing nights variable seeing conditions (seeing in the range of 0.75\arcsec -1.25\arcsec\ and 1.0\arcsec -2.5\arcsec\ for the optical and infrared data, respectively) prevented us from performing precise standard flux calibrations using spectro-photometric standards. This is because of considerable flux loss owing to the finite slit widths. To flux-calibrate the optical and infrared spectra we therefore employed the acquisition image (taken with the narrow-band filter NB\,2.191) using four field stars as ``standard stars''. Moreover, additional photometric observations of DK\,Cha were acquired eight months later using the infrared spectrograph and camera ISAAC on the ESO-VLT telescope to test the goodness of the acquisition image photometry. The ISAAC narrow-band filter at 1.21\,\um\ (where no significant spectral features were expected to contribute to the total flux)  was chosen instead of the standard J broadband filter because of the bright J-magnitude of DK\,Cha (2MASS J$\sim$9.3\,mag) that would otherwise saturate the detector. 
%The photometric standard GSPC\,S\,64-F was observed with the same instrumental configuration and similar airmass in order to calibrate the image. 
The photometric data and the acquisition image were reduced using standard IRAF \footnotemark[\value{footnote}] routines. The magnitudes derived from both the ISAAC and acquisition image photometry agree within 0.1\,mag, indicating no significant variations in the time elapsed between the observations. 
%Questa parte la toglierei:
%The spectroscopic data were then flux calibrated using the ISAAC photometry at 1.21\,\um\ taking advantage of the fact that the optical and blue spectra, as well as, the blue and red grism spectra overlap at the end and beginning of each spectral segment, respectively. This procedure insures that the line ratios are independent of flux calibrations.

\section{Results}
\label{sect:lines}

\subsection{Detected lines}

The optical and NIR spectra of DK Cha normalised to the stellar continuum are shown in Fig.\,\ref{fig:spectra}. The spectra show numerous emission lines, most of them detected at S/N greater than 5, whereas no significant absorption features are detected. Line identification together with fluxes and equivalent widths are given  in Table\,\ref{tab:lines}. 
Identification of these lines was made using the Atomic Line List database, while the goodness of the classification was checked through the detection of several transitions from the same atom and multiplet. In addition, the spectra of other YSOs \citep{kelly94} and objects with expected similar excitation conditions \citep{walmsley00} were also employed to recognise the features. 

The DK Cha spectra show permitted emission lines similar to those observed in active YSOs, as for example the presence of H$\alpha$, \HeI, the Ca II infrared triplet, and the \pab\ and \brg\ emission lines. Indeed, the most abundant features observed in our spectra are \HI\ emission lines coming from the Brackett serie.
Permitted emission lines from species such as \NaI, \MgIp, \CIp, \OIp, and \NIp\ are also detected, showing upper state excitations above 30\,000\,cm$^{-1}$, 47000\,cm$^{-1}$, 69000\,cm$^{-1}$, 88000\,cm$^{-1}$ and 100\,000\,cm$^{-1}$, respectively. Several of these lines are usually observed in the optical and infrared spectra of CTTSs and low-mass Class I sources, as well as Herbig AeBe stars, while other lines (such as the \NIp\ lines) may be excited by fluorescence, which we will discuss in Sect.\,\ref{sub:discussion-emissionlines}, and sometimes associated with photon dominated regions (PDRs; e.g. \citealt{marconi98,walmsley00}).
 
Forbidden emission lines such as \FeII, \SII\ and \OI\ transitions are also detected in our spectra. These lines are commonly accepted to be associated with jets from young stars and are often observed in the spectra of low-mass CTTSs and Class I sources showing outflow activity (see, e.g. \citealt{hartigan95, nisini_hh1,linda06}).   

Finally, in addition to the atomic lines, emission from molecular transitions are also observed in the infrared spectrum of DK Cha. In particular, weak \h\ emission at 2.12\,\um\ and band-head CO emission around 2.3\,\um\ have been detected. 
%The CO\,(2-0) transition is, however, fainter than the CO\,(3-1) and (4-2) transitions by a factor $\sim$4.4 and $\sim$3.7, respectively. A %feature in the telluric standard spectrum does not seem to be the responsible for this fact. 

\subsection{Photometry}

Table\,\ref{tab:phot} shows the photometric value retrieved from the ISAAC photometry at 1.21\,\um\ together with the V, R$_c$ and K magnitudes derived from the measured fluxes in the DK Cha spectrum at each effective wavelength. 
As already mentioned, DK Cha is known to be a variable star. \cite{hughes91} and \cite{molinari93} studied the variability of DK Cha in a five-year period (from 1987 to 1992). They reported a $\Delta$J variability up to $\sim$3\,mag in 5 years. The K- and J-magnitudes reported in the literature vary between 5-6\,mag and 9-11\,mag \citep[e.g.,][]{alcala08,hughes91}, respectively, while the V-magnitude ranges from V$\sim$18.7\,mag \citep{hughes91} to V$\sim$16.7\,mag \citep{zacharias04}.
Our measured magnitudes are within this range of values, indicating that at the time of the observations DK Cha was close to its brightness minimum.

%%%%%%%%%%%%%%%%%%%%	photometry	%%%%%%%%%%%%%%%%%%%%%%%%%%%%%%%%%%%%%
\begin{table}
\begin{minipage}[t]{\columnwidth}
\caption{DK Cha photometry.}
\label{tab:phot}
\centering
\renewcommand{\footnoterule}{}  % to avoid a line before footnotes
\begin{tabular}{c c }
\hline \hline
Filter & Magnitude  \\
\hline
 V 	& 17.9 	\\
R$_c$ 	& 16.4 \\ 
NB\_1.21& 10.5 \\
H	& 7.9 \\
 K 	& 6.0  \\
\hline
\end{tabular}
\end{minipage}
\end{table}
%%%%%%%%%%%%%%%%%%%%%%%%%%%%%%%%%%%%%%%%%%%%%%%%%%%%%%%%%%%%%%%%%%%%%%%%%%%%

\section{Analysis and discussion}
\label{sect:results}

\subsection{Extinction}
\label{sect:extinction}

Deriving the correct extinction value is crucial for the computation of YSO physical parameters. Frequently, extinction values are derived from colour excesses once a spectral type and an extinction law are assumed. 
The visual extinction (A$_V$) derived in this way depends on the adopted extinction law.
Here, we have derived A$_V$ from (I-R), (I-J) and (I-H) assuming the standard colours of \cite{kenyon95} for a F0V star \citep{spezzi08} and the reddening law of \cite{cardelli89} with R$_V$=5.5 because the interstellar medium in Chameleon clouds shows R$_V$ values around 5-6 in the densest parts of the cloud (see, e.g. \citealt{covino97,spezzi08}). The discrepancies between the A$_V$ values derived from different colours are not larger than $\sim$2\,mag (A$_V$=10.4\,mag, 10.5\,mag and 12.6\,mag for the (I-R), (I-J) and (I-H) colours), with an average value of 11$\pm$1\,mag. 
For these estimates no veiling correction to the colours was considered. 
Nevertheless, ignoring the veiling will not lead to an A$_V$ variation greater than 0.5\,mag (see \citealt{cieza05}), which is within our A$_V$ uncertainty of $\pm$1\,mag. In addition, the derived A$_V$ value 
agrees well with the extinction derived from the silicate absorption feature at 10\,\um\ as detected in the ISO-SWS spectra by \cite{acke04}. From their spectra of DK Cha, an \Av\ value ranging from $\sim$10\,mag to $\sim$12\,mag was derived, depending on whether the \citet[][\Av/$\tau(9.7)$=16.6]{rieke85} or the \citet[][\Av/$\tau(9.7)$=19.3]{mathis98} relationships were used. 
In the following we adopt an average \Av\ value of 11\,mag.

\subsection{\HI\ line emission}
\label{sect:HI}

%%%%%%%%%%%%%%%%%%%%%%%%%%%%%%%%%%%%%%%%%%%%%%%%%%%%%%%%%%%%%%%%%%%%%%%%%%%%%
\begin{figure*}
\begin{center}
 \resizebox{\textwidth}{!}{\includegraphics{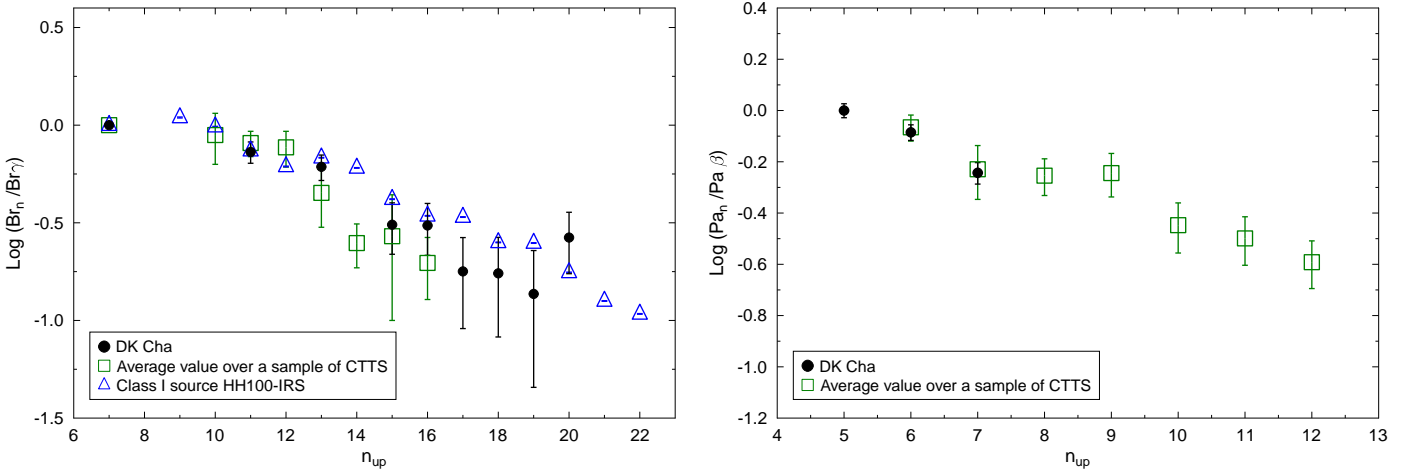}}
\end{center}
\caption{\textbf{Left panel:} Ratio of the DK Cha Brackett lines  with respect to \brg\ plotted as a function of the upper quantum number (n$_{up}$; filled black circles). For comparison the Brackett decrements for the Class I source HH100-IRS \citep[][open blue triangles]{nisini04} and the weighted mean line ratios for the Brackett series on the CTTSs sample of \citet[][open green squares]{bary08} are also included in the plot.
\textbf{Right panel:} Ratio of the DK Cha Paschen lines with respect to the \pab\ line as a function of the upper quantum number n$_{up}$ (black filled circles). For comparison, the weighted mean line ratios for the Paschen series on the CTTSs sample of \citet[][open green squares]{bary08} are also shown. }
  \label{fig:decrement1}
\end{figure*}
%%%%%%%%%%%%%%%%%%%%%%%%%%%%%%%%%%%%%%%%%%%%%%%%%%%%%%%%%%%%%%%%%%%%%%%%%%%%%

%%%%%%%%%%%%%%%%%%%%%%%%%%%%%%%%%%%%%%%%%%%%%%%%%%%%%%%%%%%%%%%%%%%%%%%%%%%%%
\begin{figure*}
\begin{center}
 \resizebox{\textwidth}{!}{\includegraphics{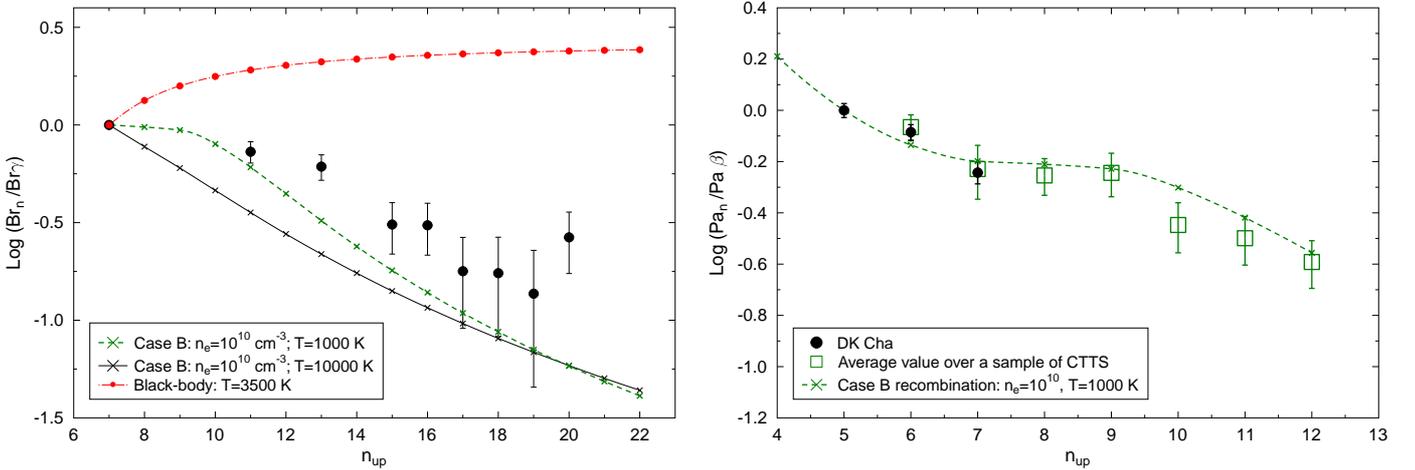}}
\end{center}
\caption{\textbf{Left panel:} Comparison between the observed DK Cha Brackett decrement and the expected curves for Case B recombination and black-body emission. The crosses represent the ratios from case B recombination at T=10\,000\,K, n=10$^{10}$\,cm$^{-3}$ (black continuous line) and T=1000\,K, n=10$^{10}$\,cm$^{-3}$ (green dashed line), while dots represent the black-body curves at T=3500\,K (dashed red line).
\textbf{Right panel:} Comparison between the observed DK Cha Paschen decrement (black dots), the Paschen decrement of the CTTSs sample of \citet[][open green squares]{bary08} and the expected Paschen ratios from Case B recombination at T=1000\,K, n=10$^{10}$\,cm$^{-3}$ (green dashed line).  }
  \label{fig:decrement2}
\end{figure*}
%%%%%%%%%%%%%%%%%%%%%%%%%%%%%%%%%%%%%%%%%%%%%%%%%%%%%%%%%%%%%%%%%%%%%%%%%%%%%

%%%%%%%%%%%%%%%%%%%%%%%%%%%%%%%%%%%%%%%%%%%%%%%%%%%%%%%%%%%%%%%%%%%%%%%%%%%%%
\begin{figure*}
\begin{center}
 \resizebox{\textwidth}{!}{\includegraphics{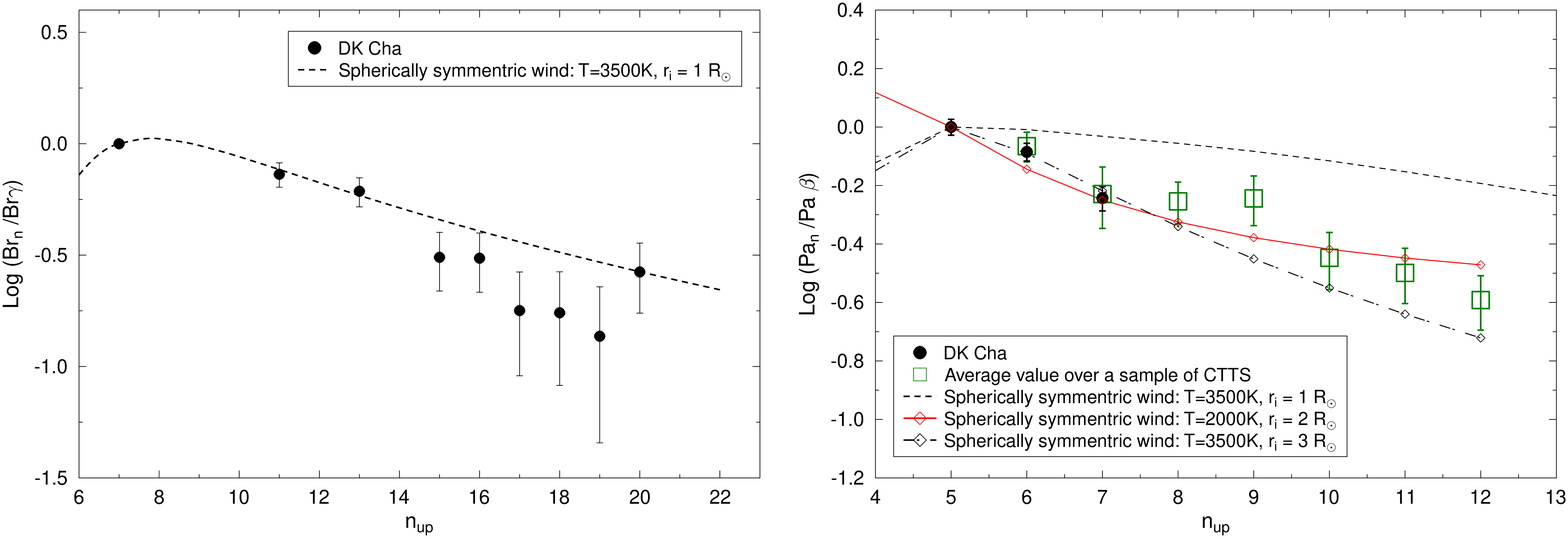}}
\end{center}
\caption{\textbf{Left panel:} Comparison between the DK Cha Brackett decrement and a model assuming an expanding wind in LTE at T=3500 K. The other parameters of the model are: r$_i$=1\,R$_\odot$, r$_{out}$=5\,R$_\odot$, n$ _{e}^{ini} $=10$ ^{13}$\cm, V$_0$=20\,\kms, V$_{max}$=300\,\kms and $\alpha$=3 (see text for details). \textbf{Right panel:} Comparison between the DK Cha Paschen decrement and the same model assuming an expanding wind in LTE at T=3500 K and r$_i$=3\,R$_\odot$ (dashed-dot black line) and at T=2000\,K and r$_i$=2\,R$_\odot$ (continuous red line). The other model parameters are the same as in the Brackett decrement case. For comparison the weighted mean line ratios for the Paschen series on the CTTSs sample of \citet[][open green squares]{bary08} were also included.}
  \label{fig:decrement3}
\end{figure*}
%%%%%%%%%%%%%%%%%%%%%%%%%%%%%%%%%%%%%%%%%%%%%%%%%%%%%%%%%%%%%%%%%%%%%%%%%%%%%

The line ratios among Brackett and Paschen lines can be used to obtain information on the physical conditions of the emitting gas, which helps us to determine the origin of the \HI\ lines. 
The SOFI spectra taken with the blue- and red-grism allow us to simultaneously detect a series of Brackett lines from \brg\ to Br\,20, and the Paschen series lines Pa$\beta$, Pa$\gamma$ and Pa$\delta$. This allows us to minimise errors in the line ratios caused by line flux variability. In addition, line ratios are independent of uncertainties in the absolute flux calibration. 

In Fig.\,\ref{fig:decrement1} the ratios of the different Brackett  and Paschen lines with respect to the \brg\ and \pab\ line (left and right panel, respectively) are shown  as a function of their upper quantum number (n$_{up}$). Only unblended lines and those in clear atmospheric windows were taken into account. The fluxes of the lines were derived from the observed equivalent widths (EWs). Because DK Cha is of spectral type F0 \citep{spezzi08}, the contribution of the intrinsic photospheric component to the observed \HI\ EWs cannot be neglected. Therefore, the EWs were corrected from the intrinsic photospheric absorption contribution following the same procedure as in \cite{rebeca06}. This procedure assumes that the observed flux is the sum of the emission from the stellar photosphere, the circumstellar gas and the disk. In addition, it assumes that the star dominates the V-band emission \citep[see,][]{rodgers01}. The latter assumption is furthermore supported by the fact that IMTTSs are not veiled around 5500\,\AA\ \citep{calvet04}. The EW of the photospheric component of each \HI\ line, corrected for the H- and K-band veiling (r$_K\sim$0.6, r$_H\sim$0.3), was computed using a F0 spectral template from \cite{rayner09}\footnote{Online at the IRTF spectral library Web site: http://irtfweb.ifa.hawaii.edu/$\sim$spex/IRTF\_Spectral\_Library/ Data\_Format.html}. No veiling was detected in the J-band. 
All relevant parameters used to derive the photospheric equivalent widths as well as the stellar parameters are given in Tables\,\ref{tab:phot} and \ref{tab:parameters}. The equivalent widths of the circumstellar \HI\ line components and the computed fluxes are shown in Table\,\ref{tab:circ}. The errors bars shown in Fig.\,\ref{fig:decrement1} correspond to an uncertainty of $\pm$1\,\AA\ in the EW$_{circ}$. In the same figure, the decrement observed in other YSOs is also plotted for comparison: namely, the line ratios as observed in the low-mass Class I source HH100-IRS \citep{linda08,nisini05} and the average values inferred from \cite{bary08} in a sample of CTTSs located in the Taurus cloud.

The DK Cha Brackett and Paschen ratios are very similar to those of low-mass CTTSs and Class I sources, although the T Tauri sample of \cite{bary08} have slightly lower Brackett ratios at N$_{up} >$ 13 with respect to DK Cha and HH100IRS. This might indicate that the excitation conditions giving rise to the Brackett series have a weak dependence on the mass and evolutionary stage of the source. 

%%%%%%%%%%%%%%%%%%%%%%% Tab Br and Pa corrette per assorbimento fotosferico %%%%%%%%%%%%%%%%%%%%%%
\begin{table}
\begin{minipage}[t]{\columnwidth}
\caption{Brackett and Paschen line fluxes.}
\label{tab:circ}
\centering
\renewcommand{\footnoterule}{}  % to avoid a line before footnotes
\begin{tabular}{c c c}
\hline \hline
Line 	    & EW$_{circ}$\footnote{Equivalent width of the circumstellar \HI\ line. Uncertainties are estimated to be $\pm$1\,\AA.}  & F\footnote{Fluxes corrected by an extinction of A$_V$=11\,mag} 			     \\
	   & (\AA)	 &  (10$^{-12}$\,erg\,s$^{-1}$\,cm$^{-2}$)	\\
\hline
Pa$\delta$  & -12.9 & 9.33 \\
Pa$\gamma$ & -18.6 & 13.4  \\
Pa$\beta$  & -23.0 & 16.3 \\
Br\,20	   & -3.0 & 2.04 \\
Br\,19	   & -1.5 & 1.02 \\
Br\,18	   & -1.9 & 1.30 \\
Br\,17	    & -2.0 & 1.33 \\
Br\,16	   & -3.4  & 2.29\\
Br\,15	   & -3.4 & 2.31 \\
Br\,13	   & -6.7 & 4.56  \\
Br\,11	   & -8.0 & 5.44 \\
Br$\gamma$ & -11.2& 7.46  \\
\hline
\end{tabular}
\end{minipage}
\end{table}

%%%%%%%%%%%%%%%%%%%%%%%%%%%%%%%%%%%%%%%%%%%%%%%%%%%%%%%%%%%%%%%%%%%%%%%%%%%%%%%%%%%%%%%%%%%%%%%%%%%

\subsubsection{Case B recombination}

To constrain the physical conditions where the Brackett and Paschen lines originate, we have explored different excitation mechanisms. In particular, \cite{bary08} found  that the Brackett and Paschen decrements of their sample of CTTSs were always fairly well fitted by a standard Case B recombination at low gas temperature ($\lesssim$2000\,K). In Fig.\,\ref{fig:decrement2} we plot the expected ratios under Case B  conditions that fit  the  Bary's sample (n=10$^{10}$\,cm$^{-3}$, and T=1000\,K), derived from the \cite{hummer87} calculations \footnote{To derive the Case B predictions, the Fortran program provided by \cite{storey95} and the required data files available at http://vizier.u-strasbg.fr/viz-bin/VizieR?-source=VI/64, have been used.}. 
We find that this model is unable to reproduce the observed Brackett decrement in DK Cha, bacause the line ratios at N$_{up}\gtrsim$12 are all underestimated. Better fits in the Case B regime are not obtained even if we increase the possible range of electron densities and temperatures: this suggests that the different behaviour of the Brackett lines in DK Cha with respect to the \cite{bary08} sample of CTTSs is not caused by the higher luminosity/mass of the source.
Unfortunately, we cannot prove which the behaviour of the Paschen lines at higher n$_{up}$ is, because only the \pab\, Pa$\gamma$ and Pa$\delta$ lines were detected at a sufficiently high S/N in a clear atmospheric region and not blended with other emission lines.

Evidence that the \HI\ lines in DK Cha cannot be reproduced by the optically thin regime assumed in the Case B recombination is also provided by the observed (extinction corrected) Pa$\beta$/Br$\gamma$ ratio. This ratio is 2.18$\pm$0.04, while for Case B it is never below 2.5: indeed, very low values of Pa$\beta$/Br$\gamma$ are obtained only when the two lines are optically thick in LTE at low temperatures \citep{gatti06,simone11}. An estimate of the temperature and projected emitting area that gives rise to this thick emission can be obtained from the Pa$\beta$/Br$\gamma$ ratio and the Pa$\beta$ flux. Assuming a line-width of 150\,\kms\ , we derive T$\sim$3500\,K and a projected area of the order of a few solar radii. This indicates that the emission originates from a quite compact region with gas at low temperature. 
To verify the consistency of our previous statement, we also plot in Fig.\,\ref{fig:decrement2} (left panel) the expected Brackett decrement for a black-body at T=3500\,K (red dotted dashed line), i.e., assuming that all Brackett lines are optically thick and at the same temperature. 
Evidently, the predicted high-n$_{up}$ lines are largely overestimated with respect to the Br$\gamma$. 
Summarising this analysis, we conclude that the DK Cha Brackett decrement is consistent with emission at a quite low temperature where the lines are optically thick for the transitions at low n$_{up}$ and optically thin for the higher n$_{up}$ values. On the other hand, no additional information can be retrieved from our Paschen decrement (Fig.\,\ref{fig:decrement2}, right panel).

As discussed by \cite{bary08}, the low temperature derived for the excitation of the Brackett and Paschen lines is not consistent with an origin in magneto-spheric accretion columns, where the temperatures are predicted to be in the range 6000-12000 K \citep{muzerolle98_MAM}.  \cite{bary08} suggested that the direct absorption of high energy photons from the  hot corona and/or accretion shock is potentially able to ionise the gas and produce the intense \HI\ emission even at low temperatures. Qualitatively  this seems applicable to the DK Cha case, which is a well known X-ray source with an X-ray luminosity of L$_X\sim$1.3$\times$10$^{31}$\,erg\,s$^{-1}$ \citep{hamaguchi05}. 

\subsubsection{Spherically symmetric expanding wind model}

Finally, to explore a different excitation scenario, we also compared the observed Brackett and Paschen decrements with the predictions of a simple model of an expanding wind, which was able  to consistently reproduce the \HI\ line emission in the Class I source HH100-IRS \citep{nisini04}. 
This model consists of a spherical envelope of ionised hydrogen in LTE , where the gas is expanding from an inner radius r$_i$ to an outer radius r$_{out}$ following a velocity law of the type V(r)=V$_0$+(V$_{max}$-V$_0$)(1-(r$_i$/r)$^\alpha$), where V$_0$ is the initial velocity at the base of the wind and V$_{max}$ is the maximum velocity of the wind \citep[see][for more details]{nisini04,nisini95}. Under these conditions, the line decrement mainly depends on the optical depth of the line, which is influenced by the velocity law, the emitting region size, and the electron density assumed at the wind base. 
Assuming a T$\sim$3500\,K and an emitting region size with radius $\sim$ 5 R$_\odot$, we are able to find quite a good agreement with all observed Brackett line ratios with a model having a V$_{max}$ $\sim$ 300 \kms\, and n$_e^{ini} \sim$ 10$^{13}$\cm\ (Fig.\,\ref{fig:decrement3}, left panel). This model is also able to reproduce the Pa$\beta$/Br$\gamma$ ratio as well as the extinction-corrected Pa$\beta$ flux of $\sim$1$\times$10$^{-11}$ erg cm$^{-2}$ s$^{-1}$.

We point out that although this is a toy model, the main results do not depend much on the given assumptions. Although the main approximation here is spherical geometry, the high gas speed and the consequent high velocity gradients imply that the radiative interaction occurs only in the confined region of line formation. Consequently, the line emissivity depends only on local physical quantities and line ratios are not much affected by geometrical effects. 

The assumption of spherical wind is instead affecting the mass loss rate implied by our model: indeed, an estimate of \.M$_w$ can be given from the continuity equation:

\begin{equation}\label{eq:mwind}
\centering
\dot{M}_W = 4\pi r_i^{2} V_{0}n_e^{ini}
\end{equation}

Substituting the relevant parameters, we derive \.M$_W$ = 3.0$\times$10$^{-7}$\,\msyr.  Such a value is of the same order of magnitude as the DK Cha mass accretion rate derived in Sect.\,\ref{sect:accretion}. However, \.M$_W$ (eq.\,\ref{eq:mwind}) scales with the total surface area and could be a small fraction of the derived value if the emission comes, e.g., from a collimated wind. 

It is also worth noting that the same wind parameters used to fit the Brackett ratios do not reproduce the Paschen decrement, however (see, Fig.\,\ref{fig:decrement3}, right panel; black dashed line). Lower temperatures (T$\sim$2000\,K) or a greater envelope thickness (around a factor of 2-3) are required to match the previous wind solution. 
This may indicate that Paschen lines are emitted in a region that is more extended than the Brackett lines and/or that a temperature gradient is required. However, our model is probably too simplistic to draw further conclusions.

\cite{kwan11} performed local excitation calculations to obtain hydrogen line opacities and emissivity ratios, finding out that the \HI\ line emission mainly arises from a highly clumpy radial outflow. Following this work, the observed Pa$\gamma$/Pa$\beta$ ratio in DK Cha (Pa$\gamma$/Pa$\beta$=0.82$\pm$0.07) is consistent with Pa$\beta$ and Pa$\gamma$ lines excited in a region with a total density of roughly N$_H\sim$1.6$\times$10$^{11}$-6$\times$10$^{11}$\,\cm\ and an optical depth $\tau_{Pa\gamma}$=33.4-210, with both lines showing different excitation temperatures in the range 3240-3920\,K and 3020-3940\,K, respectively. This result resembles that from our simple model fairly well.  

\subsection{Other detected lines.}
\label{sub:discussion-emissionlines}

Previous observations of DK Cha have already shown some of the properties of its circumstellar material. For instance, \cite{hughes91} reported the presence of a wind from this source as shown by blue-shifted forbidden line emission, whereas \cite{vankempen09} detected a CO outflow. In our spectra, the presence of faint H$_2$ emission together with the CO band-heads may also indicate the presence of a molecular outflow. However, the near-IR H$_2$ and CO emission could also be related to emission from the interaction of a wind with the surrounded medium \citep[e.g.,][]{rebeca08} or/and emission from the inner warm circumstellar disk \citep{carr93,najita96}.

The optical and NIR spectrum shows several forbidden emission lines that are typical of atomic jets, such as the several \SII\ \OI\ and \NI\ transitions between 0.6 and 1\um. The presence of a jet from DK Cha was already pointed out by \cite{hughes91}, who resolved \SII\ extended emission showing significant velocity gradients. 
From the luminosity of the \OI\,0.630\,\um\ it is possible to estimate the mass flux rate from the jet (\mjet), following the relationship given by \cite{hartigan95}. 
The main uncertainty using this expression is caused by the assumed extinction: the A$_V$ derived in Sect.\,\ref{sect:extinction} towards the central source can be considered an upper limit to the extinction towards the region of the emission of the forbidden lines, which is likely to be located farther out in a less embedded environment. Therefore, we can only give an upper limit to the mass flux by assuming A$_V$=11\,mag for the extinction towards the forbidden line emission region. In this way, we infer that \mjet\ is $\lesssim$10$^{-6}$\,\msyr, assuming a distance to the source of $\sim$178\,pc, an electron density of $\sim$10$^{5}$\cm\ \citep[because both emission from \OI\ and \SII\ have been detected, see][]{hartigan95}, and a velocity in the plane of the sky V$_\bot\sim$32\,\kms\ \citep[assuming i$\sim$18\degr\ and a radial velocity of $\sim$-98\,\kms;][]{vankempen10,hughes91}. 
To better constrain the \mjet\ value in DK Cha, the \OI\,63\,\um\ line has been used as well. This line has been detected in both ISO-LWS and HERSCHEL-PACS observations of DK Cha \citep{vankempen10,lorenzetti99}. Although this line is not affected by extinction, the \mjet\ value derived from it represents an upper limit to the total mass transported along the jet because part of the \OI\,63\,\um\ line could be excited in the PDR of the star \citep[see e.g.,][]{lorenzetti99}. Assuming a \OI\,63\,\um\ line flux of 3.7$\times$10$^{-12}$\,erg\,s$^{-1}$\,cm$^{-2}$ \citep{vankempen10} and the expression reported in \cite{cabrit02} to estimate \mjet, we found an upper limit of $\sim$3.6$\times$10$^{-7}$\,\msyr.    

HERSCHEL-PACS observations of this source also revealed the presence of several H$_2$O and OH lines interpreted as possibly generated from UV-heating of the material along the cavity walls and thus pointing out the presence of a strong UV field from the protostar \citep{vankempen10}. 
The optical and NIR spectra of DK Cha presented in this work show, indeed, several permitted transitions that are likely excited by UV pumping. 
The \OIp\,0.845\,\um\ and \OIp\,1.129\,\um\ lines are two good examples: the presence of the bright \OIp\,1.129\,\um\ emission together with the absence of the \OIp\,1.132\,\um\ line indicate \HI\ Ly\,\bet\ fluorescence of the 3d$^3$D level of \OIp\ as the pumping mechanism for the  \OIp\,0.845\,\um\ line. This suggests in turn that the H\,\alfa\ line must be optically thick in this source because a large population of the n=3 level of hydrogen is needed for this process to occur \citep{grandi75, grandi80}. Oxygen I fluorescence lines have been reported before in other Herbig AeBe stars, as MWC\,349 and R\,Mon (see e.g. \citealt{kelly94}).
In addition to the \OIp\ lines, several \NIp\ transitions at 1.013, 1.051 and 1.060\,\um, which may be also excited by fluorescence, have been observed. \cite{walmsley00} have already reported the presence of some of these lines in the spectrum of the Orion bar. They attributed the excitation of these lines to fluorescence within the ionisation front. 

\subsection{Accretion luminosity and mass accretion rate.}
\label{sect:accretion}

The accretion luminosity (\lacc) in embedded YSOs is mostly derived from the luminosity of infrared lines such as the \brg\ and \pab\ lines \citep{muzerolle98_BrG}. The \brg\ line has been successfully used to derive the accretion luminosity in objects ranging from brown-dwarfs to several solar masses \citep{calvet04,rebeca06}, however, the \pab\ line has been mostly used to compute the accretion luminosity in low-mass protostars \citep{gatti08,natta06}.
The simultaneous acquisition of the SOFI blue and red grism spectra allows us to derive \lacc\ from the luminosity of the \pab\ and \brg\ lines, which gives us the opportunity of comparing the inferred \lacc\ values without introducing any uncertainty caused by intrinsic flux line variations of the source.

With this aim, the empirical relations relating the accretion luminosity with the \brg\ and \pab\ luminosity (\citealt{calvet04} and \citealt{calvet00}) were used:

\begin{equation}
 log\, L_{acc} / L_\odot = 2.9 + 0.9\, log L(Br\gamma)
\end{equation}

\begin{equation}
  log\, L_{acc} / L_\odot = 2.80 + 1.03\, log L(Pa\beta). \\
\end{equation}

The luminosities of the \brg\ and \pab\ lines (L(\brg), L(\pab)) were derived from the observed equivalent widths of the \brg\ and \pab\ lines (EW(\brg), EW(\pab)) once they were corrected from the intrinsic photospheric absorption contribution (see Sect.\ref{sect:HI}).

Using this procedure, we found a very good agreement between the accretion luminosity computed from the \brg\ and \pab\ luminosities (see, Table\,\ref{tab:accretion}), with an average value of \lacc$\sim$9\,L$_\odot$.

Additional relationships, based on the luminosity of optical lines, have also been suggested for the determination of the accretion luminosity.
For instance, \cite{herczeg08}, \cite{dahm08} and \cite{fang09}, derived relationships based on the luminosity of the \OI, \CaIIp\ and H$\alpha$ lines, respectively. When we apply these relationships to the lines observed in DK\,Cha, we derive accretion luminosities much higher than the value estimated from the IR \HI\ lines, however (L$_{acc}\sim$104, 24 and 107\,\lsun\ for the \OI, \CaIIp\ and H$\alpha$ lines, respectively), and, more important, much higher than the bolometric luminosity of this source. The origin of this discrepancy could be related to a wrong extinction estimate. For instance, decreasing the A$_V$ of 2\,mag, new \lacc\ values of $\sim$7.4, 4.5, 11.7, 18.6 and 12\,\lsun\ are found for the \brg, \pab, \CaIIp, \OIp\ and H$\alpha$, respectively. Although these values are now lower than L$_{bol}$, the accretion luminosities derived from the \brg\ and \pab\ lines now disagree. Similar results to the ones found in DK Cha (i.e. optical lines giving rise to \lacc\ values higher than IR lines) have also been reported in a large sample of sources showing low extinction values located in the ChaI/II clouds \citep{simone11}.

It is more likely, then, that the empirical relationships found for samples of low or very low mass objects cannot be extrapolated to sources with larger mass and luminosity. In the high-luminosity objects different excitation mechanisms in addition to accretion can indeed be responsible for enhanced emission of the optical lines such as high chromosphere activity or direct excitation from stellar UV photons.  

From the accretion luminosity derived from the IR lines and stellar parameters shown in Table\,\ref{tab:parameters} a mass accretion rate (\macc) of $\sim$3.5$\times$10$^{-7}$\,\msyr\ was found from the expression \citep{gullbring98} 
\begin{equation}
\dot{M}_{acc} = \frac{L_{acc} R_*}{G M_*} \left(1-\frac{R_*}{R_{in}}\right)^{-1},
\end{equation}
where R$_*$ and M$_*$ are the stellar radius and mass, G is the universal constant and a R$_{in}$=5\,R$_{\odot}$ is the inner disk radius. The inferred \macc\ value is within the range of those found in classical Herbig Ae stars and intermediate mass T Tauri stars (IMTTS) where \macc\ in the range 10$^{-6}$-10$^{-8}$\,\msyr\ are typically found \citep{donehew11,rebeca06, calvet04,rodgers01}. 
Interestingly, when one compares sources of roughly the same mass, the \macc\ value found in DK Cha is higher than that found in the IMTTS sample of \cite{calvet04}  and in \cite{rebeca06} (where \macc\ values $\sim$10$^{-8}$\,\msyr\ were found) but lower than that computed for the Class I sample of \cite{white04} (\macc\ $\sim$10$^{-6}$\,\msyr). This result is consistent with DK Cha being in an evolutionary transition phase between a Class I and II source.

\begin{table}
\begin{minipage}[t]{\columnwidth}
\caption{DKCha stellar parameters.}
\label{tab:parameters}
\centering
\renewcommand{\footnoterule}{}  % to avoid a line before footnotes
\begin{tabular}{c c c c c c c }
\hline \hline
ST & D   &  T$_{eff}$ & L$_*$ & L$_{bol}$ & R$_*$ & M$_*$ \\
   & (pc)& (K)       & (L$_{\odot}$) & (L$_{\odot}$) & (R$_{\odot}$) & (M$_{\odot}$) \\
\hline
F0\footnote{From \cite{spezzi08}} & 178  &  7200$^a$		 & 18.62$^a$ &   29.4\footnote{From \cite{vankempen10}}  & 2.77$^a$  & 2$^a$     \\
\hline
\end{tabular}
\end{minipage}
\end{table}
%%%%%%%%%%%%%%%%%%%%%%%%%%%%%%%%%%%%%%%%%%%%%%%%%%%%%%%%%%%%%%%%%%%%%%%%%%%%

%%%%%%%%%%%%%%%%%%%%	EWs	%%%%%%%%%%%%%%%%%%%%%%%%%%%%%%%%%%%%%
%\begin{table}
%\begin{minipage}[t]{\columnwidth}
%\caption{\pab\  and \brg\ equivalent widths.}
%\label{tab:ew}
%\centering
%\renewcommand{\footnoterule}{}  % to avoid a line before footnotes
%\begin{tabular}{ c  c c }
%\hline \hline
%Line &	EW$_{obs}$ & EW$_{circ}$	 \\
%     &	(\AA)	   &  (\AA)      \\ 
%\hline                        
%
%\pab\ & -17.3 & -23.0 \\
%\brg\ & -8.7 & -11.2 \\
%
%\hline
%\end{tabular}
%\end{minipage}
%\end{table}
%%%%%%%%%%%%%%%%%%%%%%%%%%%%%%%%%%%%%%%%%%%%%%%%%%%%%%%%%%%%%%%%%%%%%%%%%%%%

%%%%%%%%%%%%%%%%%%%%	Lacc	%%%%%%%%%%%%%%%%%%%%%%%%%%%%%%%%%%%%%
\begin{table}
\begin{minipage}[t]{\columnwidth}
\caption{Accretion properties.}
\label{tab:accretion}
\centering
\renewcommand{\footnoterule}{}  % to avoid a line before footnotes
\begin{tabular}{c |  c c | c }
\hline \hline
A$_V$  & \lacc(\pab) & \lacc(\brg)	& \macc	 \\
(mag) &	\multicolumn{2}{c}{(L$_{\odot}$)} & (10$^{-7}$\,\msyr)\\
\hline                        

11.2 & 9.5 & 9.1 & 3.5  \\

\hline
\end{tabular}
\end{minipage}
\end{table}
%%%%%%%%%%%%%%%%%%%%%%%%%%%%%%%%%%%%%%%%%%%%%%%%%%%%%%%%%%%%%%%%%%%%%%%%%%%%

\section{Conclusions}
\label{sect:conclusions}

We have presented low-resolution spectroscopic observations of the embedded Herbig Ae star DK Cha covering the range 0.6\,\um-2.4\,\um.  Its nearly face-on configuration has allowed us to have direct access to the star-disk system 
and to detect numerous emission lines tracing different emission regions and excitation conditions in the nearby surroundings of the protostar. The extremely rich emission line spectrum makes this source
very similar to low-mass very active T Tauri stars. We summarise our results as follows:

\begin{itemize}
 \item The DK Cha spectrum shows several permitted (e.g., \CaIIp,\MgIp,\NaI), forbidden (e.g.,\FeII, \SII) and molecular emission lines (CO and H$_2$). The \HI\ lines are those more numerous along the spectrum. In addition, some of the permitted lines, such as the \OIp\,0.845\,\um\ line, were identified as being excited by fluorescence.

%\item From the color excesses a visual extinction of 11.2$\pm$1\,mag was found.

\item To constrain the origin of the \HI\ lines, Brackett decrement plots were constructed and compared with different excitation mechanisms. Case B recombination is not able to fit all observed Brackett ratios, while optically thick emission in LTE at a temperature of $\sim$ 3500 K is required to account for the observed \pab/\brg\, ratio. A simple model for an expanding gas is able to reproduce all observed Brackett ratios by assuming that this gas is located very close to the central source and has a high density at its base, of the order of 10$^{13}$\cm. The same parameters used to reproduce the Brackett ratios cannot account for the Paschen line ratios.

\item From the \pab\ and \brg\ line luminosity we derived the accretion luminosity using the expressions from \cite{calvet00,calvet04}. We found a very good agreement between the \lacc\ derived from both expressions, with an average value of \lacc$\sim$9\,\lsun. From the accretion luminosity a \macc$\sim$3.5$\times$10$^{-7}$\,\msyr\ was found. The measured \macc\ value is consistent with DK Cha being in an evolutionary transition phase between a Class I and II source. 
%of the same order as that found in older Herbig Ae and IMTTSs. This may indicate that DK Cha is in a final stage of accretion, with an L$_{acc}$/L$_{bol}$ less than 0.5, or that most of the matter is accumulated in outbursts of accretion   that could be responsible for the high variability of the object.

\end{itemize}

\begin{acknowledgements}
The authors kindly thanks the following people (in alphabetical order) who contributed to the Atomic Line List v2.05b12 (http://www.pa.uky.edu/$\sim$peter/atomic/) by providing data and/or helpful insights: K.M. Aggarwal, M.A. Bautista, R. Kisielius, S.N. Nahar, M.J. Seaton, D.A. Verner. We are grateful to Malcolm Walmsley and Antonella Natta for fruitful discussions and insightful comments. RGL, ACG and TR were supported by the Science Foundation Ireland, grant 07/RFP/PHYF790. ACG also acknowlegdes support from the European Commision, grant ERG249157. 
\end{acknowledgements}

\bibliographystyle{aa}
\bibliography{references}

\end{document}